\documentclass[a4paper,11pt]{article}
\usepackage{jheppub}
\usepackage{graphicx}
\usepackage{amsmath}
\usepackage{amsfonts}
\usepackage{amssymb}
\usepackage{slashed}
\usepackage[colorlinks=true,linkcolor=blue]{hyperref}
\usepackage{bm}
\usepackage{empheq}
\usepackage[super]{nth}
\def\sss{\scriptscriptstyle}

\def\tA{\theta_{\!\sss{A}}}

\def\tstart{t_{\mathrm{start}}}
\def\kstart{k_{\mathrm{start}}}

\def\rmin{r_{\mathrm{min}}}

\def\OO{\mathcal{O}}
\def\LL{\mathcal{L}}
\def\be{\begin{equation}}
\def\ee{\end{equation}}
\def\bea{\begin{eqnarray}}
\def\eea{\end{eqnarray}}

\def\Eq#1{Eq.~\eqref{#1}}

\title{How to simulate global cosmic strings with large string tension}
\author{Vincent B.\ Klaer, Guy D.\ Moore}

\affiliation{Institut f\"ur Kernphysik, Technische Universit\"at Darmstadt\\
Schlossgartenstra{\ss}e 2, D-64289 Darmstadt, Germany}
\emailAdd{vklaer@theorie.ikp.physik.tu-darmstadt.de,guy.moore@physik.tu-darmstadt.de}

\abstract{
Global string networks may be relevant in axion production in the
early Universe, as well as other cosmological scenarios.  Such networks
contain a large hierarchy of scales between the string core scale and
the Hubble scale, $\ln(f_a/H) \sim 70$, which influences the network
dynamics by giving the strings large tensions $T \simeq \pi f_a^2
\ln(f_a/H)$.  We present a new numerical approach to simulate such
global string networks, capturing the tension without an exponentially
large lattice.
}

\begin{document}
\maketitle
\section{Introduction}
\label{sec:intro}

Cosmic strings are extended solitonic objects which arise in field
theories when a symmetry $G$ is broken, $G \to H$, and the quotient
space $G/H$ has nontrivial $\pi_1$ homotopy
\cite{Kibble:1976sj,Vilenkin:1984ib,Gibbons:1990gp}.
The simplest example is when a complex scalar field
$\varphi = \frac{\varphi_r + i \varphi_i}{\sqrt 2}$ possesses a U(1)
symmetry, $\varphi \to e^{i\tA} \varphi$, but the potential leads to
spontaneous symmetry breaking,%
\footnote{We use natural units and $[{-}{+}{+}{+}]$ metric signature.}
\begin{equation}
  \label{eq:L}
- \LL = \partial_\mu \varphi^* \partial^\mu \varphi
+ \frac{m^2}{8v^2} \Big( 2 \varphi^* \varphi - f_a^2 \Big)^2 \,.
\end{equation}
Here $m$ is the mass of the radial (Higgs) excitation and $f_a$ is the
vacuum value of the scalar field.
Under this Lagrangian, the vacuum state is of form
$\varphi = \sqrt{2} f_a e^{i\tA}$, and the arbitrary angle $\tA$
breaks the U(1) symmetry completely.  If the field initially makes
this choice of symmetry breaking direction randomly and independently
at widely separated points in space, then the initial
conditions generically contain string type defects
(the Kibble mechanism \cite{Kibble:1976sj}).  Qualitatively, the same
lessons are true if $\varphi$ is charged under a U(1) gauge symmetry.
We will refer to the case with gauge symmetry as a theory of local or
abelian strings, and the theory where the U(1) symmetry is a global
symmetry as a theory of global or scalar-only strings.

If a network of such defects exists in the Universe, it could
influence the development of cosmic structure
\cite{Zeldovich:1980gh,Vilenkin:1981iu,Kibble:1980mv}.
A more physically interesting scenario, in light of strong limits on
the role of strings in cosmic structure
\cite{Ade:2013xla,Urrestilla:2011gr,Lizarraga:2014xza,Lazanu:2014xxa},
is the possibility that the QCD axion
\cite{Weinberg:1977ma,Wilczek:1977pj}
makes up the Dark Matter of the Universe
\cite{Preskill:1982cy,Abbott:1982af,Dine:1982ah}.
The QCD axion's Lagrangian is the same as \Eq{eq:L}, with the addition
of a small temperature-dependent explicit symmetry breaking
term $\chi(T) ( \cos[\mathrm{Arg}\,\varphi] - 1)$,
which becomes important around the QCD scale.  If the axion
field starts out with randomly different values for the
symmetry-breaking angle $\tA$ at different space points, which is
plausible and maybe even likely \cite{Visinelli:2014twa},
then there would be an initial network of (axionic) global cosmic
strings.  This network is destroyed around the QCD scale when the
potential's tilt becomes important, but it could play a dominant role
in establishing the density of axions produced
\cite{Davis:1986xc}.

If we could understand the efficiency of axion production in the early
Universe, then assuming the axion makes up the dark matter, we could
predict the axion mass \cite{Visinelli:2014twa}.
The problem is that we don't understand the string network evolution
well enough.  There is no consensus in the literature for the
efficiency of axion production
\cite{Davis:1989nj,Dabholkar:1989ju,Hagmann:1990mj,%
Battye:1993jv,Battye:1995hw,Chang:1998tb,%
Yamaguchi:1998iv,Yamaguchi:1998gx,Yamaguchi:1999yp,Yamaguchi:1999dy,%
Hagmann:2000ja,Martins:2003vd,Wantz:2009it,%
Hiramatsu:2010yn,Hiramatsu:2010yu,Hiramatsu:2012gg,%
Hiramatsu:2012sc,Kawasaki:2014sqa}.
Recent large-scale numerical simulations
\cite{Kawasaki:2014sqa,axion1}
have not resolved this problem, because no simulation to date can
correctly treat the tension of a global string.  As we will now
explain, numerical simulations of global string networks typically
study networks where the strings have a tension $\OO(10)$ times
smaller than the physically relevant value.  This \textsl{may} be
dramatically misrepresenting the density, longevity, and role of the
strings in these simulations, and therefore the amount of axions
produced \cite{axion1}.

To understand the problem, consider the field solution for a string
under the Lagrangian of \Eq{eq:L}.  Choosing the string to lie along
the $z$ axis in polar coordinates $(z,r,\phi)$, the string solution is
\begin{align}
  \label{string_sol}
  \varphi(z,r,\phi) & = e^{i\phi} \,f(mr)\:  f_a \sqrt{2} \,, \\
  \label{string2}
  f''(x) & = -\frac{x f'(x) + f(x)}{x^2} - \frac{f(x)(1-f(x)^2)}{2}
  \,, \\
  f(0) & = 0 , \qquad \lim_{x\to \infty} f(x) = 1 \,. \nonumber
\end{align}
Here we have shown the differential equation and boundary values which
the radial function $f(mr)$ should obey.  This solution should be
valid out to a value of $r$ of order the distance to the next string,
which is parametrically $Hr \sim 1$ (with $H$ the Hubble parameter).
In terms of the parameter $x=mr$, this is a value of $x \sim m/H \gg 1$.
For instance, for an axion near the QCD scale,
$H \sim T^2/m_{\mathrm{pl}} \sim 10^{-19} \, \mathrm{GeV}$
while $m \sim f_a \sim 10^{11} \, \mathrm{GeV}$ (for a typical estimate
of $f_a$ \cite{diCortona:2015ldu}).
To see why this is a problem for simulations, we estimate the energy
per length, or tension $T$, stored in the string:
\begin{align}
  \label{tension}
  T & = \int_0^{2\pi} d\phi \int_0^{\sim H^{-1}} r \: dr \left[
    \frac{1}{2} |\partial_r \varphi|^2 +
    \frac{1}{2r^2} |\partial_\phi \varphi|^2 \right]
  \\ & \simeq 2\pi \int_{\sim m^{-1}}^{\sim H^{-1}} r\:dr
  \; \frac{1}{2r^2} f_a^2 \simeq \pi f_a^2 \ln \frac{m}{H}
  \equiv \pi f_a^2 \kappa \,. \nonumber
\end{align}
The string tension is logarithmically small-$r$ divergent,
because the derivative in the $\phi$ direction is proportional to
$1/r$.  Therefore the tension contains a logarithmically large factor
$\ln(m/H)$, which for the values quoted above is
$\ln(10^{30}) \simeq 70$.  If the goal is to study global strings
playing a role in structure formation in the modern Universe, and they
arise from a GUT or other high-scale theory, then the logarithm
may be more like
$\ln(10^{15}\:\mbox{GeV}\times 10^{10} \: \mbox{year}) \sim 113$.
We have named the magnitude of this logarithm $\kappa$.

On the other hand, existing numerical simulations of global string
networks rely on modifying \Eq{eq:L} to incorporate Hubble expansion,
in comoving coordinates and conformal time, and solving it
numerically as a function of (conformal) time on a
spatial lattice.  To properly treat the string defects, it is
necessary that the lattice \textsl{resolves} the string cores, that
is, the lattice spacing $a$ must satisfy $ma \lesssim 1$.
At the same time, the lattice box-size must be larger than the typical
inter-string spacing, and is typically larger by a factor of at least
a few.  Therefore, in the numerical simulation, the ratio of string
separation-to-core is constrained to be at most a few hundred, and the
logarithm of interest is at most 5 or 6.
In other words, in existing simulations
\cite{Hiramatsu:2012gg,axion1}, the global strings have a string tension
at least an order of magnitude too small.

Does this matter?  Probably it does.  Global string networks differ
from the much better-studied local string networks
\cite{Albrecht:1989mk,Dabholkar:1989ju,Bennett:1989yp,%
Allen:1990tv,Vanchurin:2005yb,Olum:2006ix}
in two important ways:
\begin{enumerate}
\item
  \label{effect1}
  There is a massless field coupled to the string, namely the
  Goldstone boson mode $\tA$ (the phase of $\varphi$).  The strings
  can radiate away their energy to this field.  The strength of the
  interaction is governed by $f_a^2$.  Radiation should make it easier
  for strings to lose energy, leading to a less-dense network of
  smoother strings.
\item
  \label{effect2}
  The massless field also communicates inter-string forces.  These
  could help the strings to find each other and annihilate, again
  leading to a smaller string density.  The strength of the
  interactions is again governed by $f_a^2$.
\end{enumerate}
These effects compete against string-tension effects which scale
as $T = \pi \kappa f_a^2$.  Therefore the global-string effects are
suppressed by a factor of $1/\kappa$, and are $\sim 10$ times
\textsl{less} important in true string network evolutions than in
those which we can simulate.

We know that these effects play a major role in string network
evolution, because the density of the string network in global string
simulations is about a factor of 4--8 smaller than in local string
network simulations
\cite{Yamaguchi:1998gx,Bennett:1989yp,%
Allen:1990tv,Vanchurin:2005yb,Olum:2006ix,Hindmarsh:2017qff}.
Therefore, the correct inclusion of the high tension of string cores
may lead to a substantial change in the string network dynamics and
the network density.  Indeed, in the \textsl{limit}
$\kappa \gg 1$, we expect the global string networks to become
indistinguishable from local strings.  Probably $\kappa=70$ is not
enough to achieve this limit, but it should certainly give different
string dynamics than $\kappa = 5$.
In the context of axion production, the denser network with a larger
string tension means there is more energy available for the
production of axions.  But the large tension makes the strings more
robust to external forces, so the network should be more persistent
once the potential tilts, and will take longer to annihilate away.

It is difficult to extrapolate the consequences of the high string
tension on axion production \cite{Kawasaki:2014sqa}.
Therefore, a method to simulate global string networks with $\kappa
\sim 70$ is clearly well motivated.  One of us recently presented such
a method and implemented it in 2 space dimensions \cite{axion2}.  The
results indicated that the produced axion density rises rather
modestly with string tension.  However, 2
dimensions may show quite different physics than 3, and we have not
(yet) been able to extend the method proposed in \cite{axion2} to 3
space dimensions.  Instead, in this paper we will propose another
approach to simulate a global string network with enhanced string core
tension.  Essentially, we will present a model in which each global
string has an abelian-Higgs string bound onto its core.  The abelian
Higgs string provides most of the string tension, and the long-range
interactions are controlled by the global Goldstone fields.  The ratio
of tensions is tunable and can be chosen to make $\kappa \sim 70$
without much difficulty.  The next section, Section
\ref{sec:method}, explains the model which does this and justifies
that the relevant infrared physics should be correct.
We study the resulting string networks numerically in
Section \ref{numerics}, and present a discussion and conclusions
in Section \ref{discussion}.  Very briefly, we find that as $\kappa$
is increased, the resulting string network grows denser and its
properties (velocity, cuspiness) change from those of a global network
towards those of an abelian-Higgs network.  Applications to axion
production are postponed to a follow-up paper.

\section{How to get large string tensions in a global model}
\label{sec:method}

We are interested in the large-scale structure of string
networks and the infrared behavior of any (pseudo)Goldstone modes they
radiate.  For these purposes it is
not necessary to keep track of all physics down to the scale of the
string core.  Rather, it is sufficient to describe the desired IR behavior
with an \textsl{effective theory} of the strings and the Goldstone
modes around them.  This consists of removing the physics very close
to the string core with an equivalent set of physics.  It has long
been known how to do this \cite{Dabholkar:1989ju}.  The string cores
are described by the Nambu-Goto action
\cite{Goto:1971ce,Goddard:1973qh,Nambu:1974zg}, which describes the
physics generated by the string tension arising close to the string
core.  The physics of the Goldstone mode is described by a Lagrangian
containing the scalar field's phase.  And they are coupled by the
Kalb-Ramond action \cite{Kalb:1974yc,Vilenkin:1986ku}:
\begin{align}
  \label{Ltot}
  \LL & = \LL_{\mathrm{NG}} + \LL_{\mathrm{GS}} + \LL_{\mathrm{KR}} \,,
  \\
  \label{LNG}
  \LL_{\mathrm{NG}} & =  \kappa \pi f_a^2 \int d\sigma
  \sqrt{{y'}^2(\sigma)(1-\dot{y}^2(\sigma))} \,,
  \\
  \label{LGS}
  \LL_{\mathrm{GS}} & = f_a^2 \int d^3 x \; \partial_\mu \tA \partial^\mu \tA \,,
  \\
  \label{LKR}
  \LL_{\mathrm{KR}} & = \int d^3 x \; A_{\mu\nu} j^{\mu\nu} \,,
  \\
  \label{Hmna}
  H_{\mu\nu\alpha} & = f_a \epsilon_{\mu\nu\alpha\beta}
  \partial^\beta \tA = \partial_\mu A_{\nu\alpha} + \mbox{cyclic} \,,
  \\
  \label{jmn}
  j^{\mu\nu} & = -2\pi f_a \int d\sigma
  \left( v^\mu {y'}^\nu - v^\nu {y'}^\mu \right) \delta^3
  (x-y(\sigma))
  \,.
\end{align}
Here $\sigma$ is an affine parameter describing the string's location
$y^\mu(\sigma,t)$, $v^\mu=(1,\dot y)=dy^\mu/dt$ is the string velocity, and
$H_{\mu\nu\alpha}$ and $A_{\mu\nu}$ are the Kalb-Ramond field strength
and tensor potential, which are a dual representation of $\tA$.
Effectively $\LL_{\mathrm{NG}}$ tracks the effects of the string
tension, which we name $\kappa \pi f_a^2$, stored locally along its
length.  Next, $\LL_{\mathrm{GS}}$ says that the axion angle
propagates under a free wave
equation, as expected for a Goldstone boson, and its decay constant is
$f_a$.  And $\LL_{\mathrm{KR}}$ incorporates the interaction between
strings and axions, also controlled by $f_a$.  The interaction can be
summarized by saying that the string forces $\tA$ to wind by $2\pi$ in
going around the string (in the same sense that the $eJ_\mu A^\mu$
interaction in electrodynamics can be summarized by saying that it
enforces that the electric flux emerging from a charge is $e$).

It should be emphasized that in writing these equations, we are
implicitly assuming a separation scale $\rmin$; at larger
distances from a string $r>\rmin$ we consider $\nabla
\varphi$ energy to be associated with $\tA$; for $r <
\rmin$ the gradient energy is considered as part of the
string tension \cite{Dabholkar:1989ju}.

Any other set of UV physics which reduces to the effective description
of \Eq{Ltot} would present an equally valid way to study this string
network.  Our plan is to find a model without a large scale hierarchy,
such that the IR behavior is also described by \Eq{Ltot} with a large
value for the string tension. Optimally, we want a model which is easy
to simulate on the lattice with a spacing not much smaller than
$\rmin$.  Reading \Eq{LNG} through \Eq{jmn} in order, the
model must have Goldstone bosons with a decay constant $f_a$ and
strings with a large and tunable tension
$T = \kappa \pi f_a^2$, with $\kappa \gg 1$.  There can be other 
degrees of freedom, but only if they are very heavy (with mass
$m \sim \rmin^{-1}$), and we will be interested in the
limit that their mass goes to infinity.  Finally, the string must have
the correct Kalb-Ramond charge.  Provided everything is derived from
an action, this will be true if the Goldstone boson mode always winds
by $2\pi$ around a loop which circles a string.

We do this by writing down a model of \textsl{two} scalar fields
$\varphi_1,\varphi_2$, each with a U(1) phase symmetry.  A linear
combination of the phases is gauged; specifically, the fields are
given electrical charges $q_1 \in \mathcal{Z}$ and $q_2 = q_1-1$ under
a single U(1) gauge field.
The orthogonal phase combination represents a global U(1) symmetry
which will give rise to our Goldstone bosons.  The role of the gauge
symmetry will be to attach an abelian-Higgs string onto every global
string, which will enhance the string tension.  The added degrees of
freedom are all massive off the string, achieving our intended
effective description.

Specifically, the Lagrangian is
\begin{align}
  \label{L-2field}
  - \LL(\varphi_1,\varphi_2,A_\mu) & =
  \frac{1}{4e^2} F_{\mu\nu} F^{\mu\nu}
  + \Big| (\partial_\mu -i q_1 A_\mu) \varphi_1 \Big|^2
  + \Big| (\partial_\mu -i q_2 A_\mu) \varphi_2 \Big|^2
  \nonumber \\ & \phantom{=} {} +
  \frac{m_1^2}{8 v_1^2} \Big( 2\varphi_1^* \varphi_1 - v_1^2 \Big)^2
  + \frac{m_2^2}{8 v_2^2} \Big( 2\varphi_2^* \varphi_2 - v_2^2 \Big)^2
  \nonumber \\ & \phantom{=} {}
  + \frac{\lambda_{12}}{2} \Big(  2\varphi_1^* \varphi_1 - v_1^2 \Big)
  \Big( 2\varphi_2^* \varphi_2 - v_2^2 \Big) \,.
\end{align}
For simplicity we will specialize to the case
\begin{equation}
  \label{masses_equal}
  \lambda_{12}=0, \quad m_1=m_2=\sqrt{e^2(q_1^2 v_1^2 + q_2^2 v_2^2)}
  \equiv m_e \,.
\end{equation}
The model has 6 degrees of freedom; two from each scalar and two from
the gauge boson.  Symmetry breaking,
$\varphi_1 = e^{i\theta_1} v_1\sqrt{2}$ and
$\varphi_2 = e^{i\theta_2} v_2 \sqrt{2}$, spontaneously breaks both
U(1) symmetries and leaves five massive and one massless degrees of
freedom. Specifically, expanding about a vacuum configuration, the
fluctuations and their masses are
\begin{align}
  \label{m-h1}
  v_1 & \to v_1 + h_1 \,, && m = m_1 \\
  \label{m-h2}
  v_2 & \to v_2 + h_2 \,, && m = m_2 \\
  \label{m-A}
  A_i  & \neq 0 \,, && m = \sqrt{e^2(q_1^2 v_1^2 + q_2^2 v_2^2)}\equiv m_e \\
  \label{m-eaten}
  (\theta_1,\theta_2) & \to (\theta_1,\theta_2) + \omega (q_1,q_2) \,, &&
  \mbox{eaten by $A$} \\
  \label{m-goldstone}
  (\theta_1,\theta_2) & \to (\theta_1,\theta_2) +
  \tA \left( \textstyle{ \frac{q_2}{{q_1^2+q_2^2}} ,
    \frac{-q_1}{{q_1^2 + q_2^2}}} \right) && m = 0 \,.
\end{align}
We see that the choices in \Eq{masses_equal} have made all heavy
masses equal.%
\footnote{%
  We set $\lambda_{12}=0$ so that the fluctuations
  in $|\varphi_1|$ and $|\varphi_2|$ are unmixed; our other choices
  ensure that all heavy fields have the same mass.  We could consider
  other cases but we see no advantage in doing so if the goal is to
  implement the model on the lattice.  The lattice spacing is limited
  by the inverse of the heaviest particle mass, while the size of the
  string core and the mass of extra degrees of freedom off the string
  will be set by the inverse of the lightest particle
  mass.  So we get a good continuum limit with the thinnest strings,
  and therefore the best resolution of the network, by having all
  heavy masses equal.}
To clarify, note that a gauge transformation
$A_\mu \to A_\mu + \partial_\mu \omega$ changes
$\theta_1 \to \theta_1 + q_1 \omega$ and
$\theta_2 \to \theta_2 + q_2 \omega$.  Therefore the linear
combination of $\theta_1,\theta_2$ fluctuations with
$\delta \theta_1 \propto q_1$ and $\delta \theta_2 \propto q_2$ is 
precisely the combination which can be shifted into $A^\mu$ by a gauge
change, and is therefore the combination which is ``eaten'' by the
$A$-field to become the third massive degree of freedom.  The
remaining phase difference $q_2 \theta_1 - q_1 \theta_2$ is gauge
invariant,
\begin{equation}
  \label{gauge-invariant}
  q_2 \theta_1 - q_1 \theta_2 \to_{\omega} q_2(\theta_1 + q_1 \omega)
  - q_1(\theta_2 + q_2 \omega)
  = q_2 \theta_1 - q_1 \theta_2 + 0 \omega
\end{equation}
and represents a global, Goldstone-boson mode.

The model breaks two U(1) symmetries, so we must describe strings in
terms of a double $(m,n)$ representing the winding number of each
scalar field around the string.  In the absence of gauge interactions
(for $e\to 0$) the $\varphi_1,\varphi_2$ fields would not interact and
the tension of the $(1,1)$ string would be the sum of the tensions of
a $(1,0)$ and a $(0,1)$ string.  Therefore the $(1,1)$ string would be
neutrally stable to splitting into $(1,0)$ and $(0,1)$ strings.
Gauge interactions strongly change this, such that $(1,0)$ and $(0,1)$
strings strongly attract and $(1,1)$ strings are stable.  To show this
we analyze the form of a $(j,k)$ string.  For the \textsl{Ansatz}
\begin{align}
  \label{Ansatz}
  \sqrt{2} \varphi_1(r,\phi) & = e^{ij \varphi} f_1(r) v_1 \,, \nonumber \\
  \sqrt{2} \varphi_2(r,\phi) & = e^{ik \varphi} f_2(r) v_2 \,, \nonumber \\
  A_\phi(r) & = \frac{g(r)}{r} \,,
\end{align}
we find the equations of motion from \Eq{L-2field} are
\begin{align}
  \label{g-eq}
  g'' - \frac{g'}{r} & = e^2 v_1^2 f_1^2 q_1(q_1 g-j)
  + e^2 v_2^2 f_2^2 q_2(q_2 g-k) \,, \\
  \label{f1-eq}
  f_1'' + \frac{f_1'}{r} & = \frac{f_1}{r^2} (j-q_1 g)^2
  + \frac{m^2}{2} f_1(f_1^2-1) \,, \\
  \label{f2-eq}
  f_2'' + \frac{f_2'}{r} & = \frac{f_2}{r^2} (k-q_2 g)^2
  + \frac{m^2}{2} f_2(f_2^2-1) \,.
\end{align}
Here $f_1,f_2$ represent the progress of the two scalar fields towards
their large-radius asymptotic vacuum values, while $2\pi g(r)$ is the
magnetic flux enclosed by a loop at radius $r$, which trends at large
$r$ towards the total enclosed magnetic flux.
The large-$r$ behavior is well behaved only if $f_1\to 1$,
$f_2 \to 1$, and
\begin{align}
  \label{g-limit}
  \lim_{r\to \infty}  g(r) = \frac{j q_1 v_1^2 + k q_2 v_2^2}
      {q_1^2 v_1^2 + q_2^2 v_2^2}
      \quad = \frac{1}{2\pi}\mbox{(enclosed magnetic flux)}\,.
\end{align}
The magnetic flux is therefore a compromise between the value
$j/q_1$, which cancels large-distance gradient energies for the first
field, and $k/q_2$, which cancels large-distance gradient energies for
the second field.  Unless $q_2 j - q_1 k=0$, the gradient energies are
not canceled at long distance.  Indeed, we can understand the
difference $(q_2 j - q_1 k)$ as the global (axionic) charge of the
string.  The gradient energy at large distance is given by
\begin{align}
  \label{tension-fa}
  T & \simeq 2\pi \int r \: dr \left( |D_\phi \varphi_1|^2 +
  |D_\phi \varphi_2|^2 \right) \nonumber \\
  & \simeq \pi \int r \: dr \left( \frac{v_1^2}{r^2}
  \Big( j - q_1 g \Big)^2 + \frac{v_2^2}{r^2}
  \Big( k - q_2 g \Big)^2 \right)
  \nonumber \\
  & \simeq \pi \int \frac{dr}{r} \:
  \frac{v_1^2 v_2^2 (jq_2-kq_1)^2}{q_1^2 v_1^2 + q_2^2 v_2^2} \,,
\end{align}
which is proportional to the squared global charge of the string.
Naming $q_{\mathrm{global}} = jq_2 - kq_1$, comparing
\Eq{tension} with \Eq{tension-fa}, we identify the Goldstone-mode
decay constant as
\begin{equation}
  \label{fa-is}
  f_a^2 = \frac{v_1^2 v_2^2}{q_1^2 v_1^2 + q_2^2 v_2^2} \,.
\end{equation}

\begin{figure}[htb]
  \hfill \epsfxsize=0.4\textwidth\epsfbox{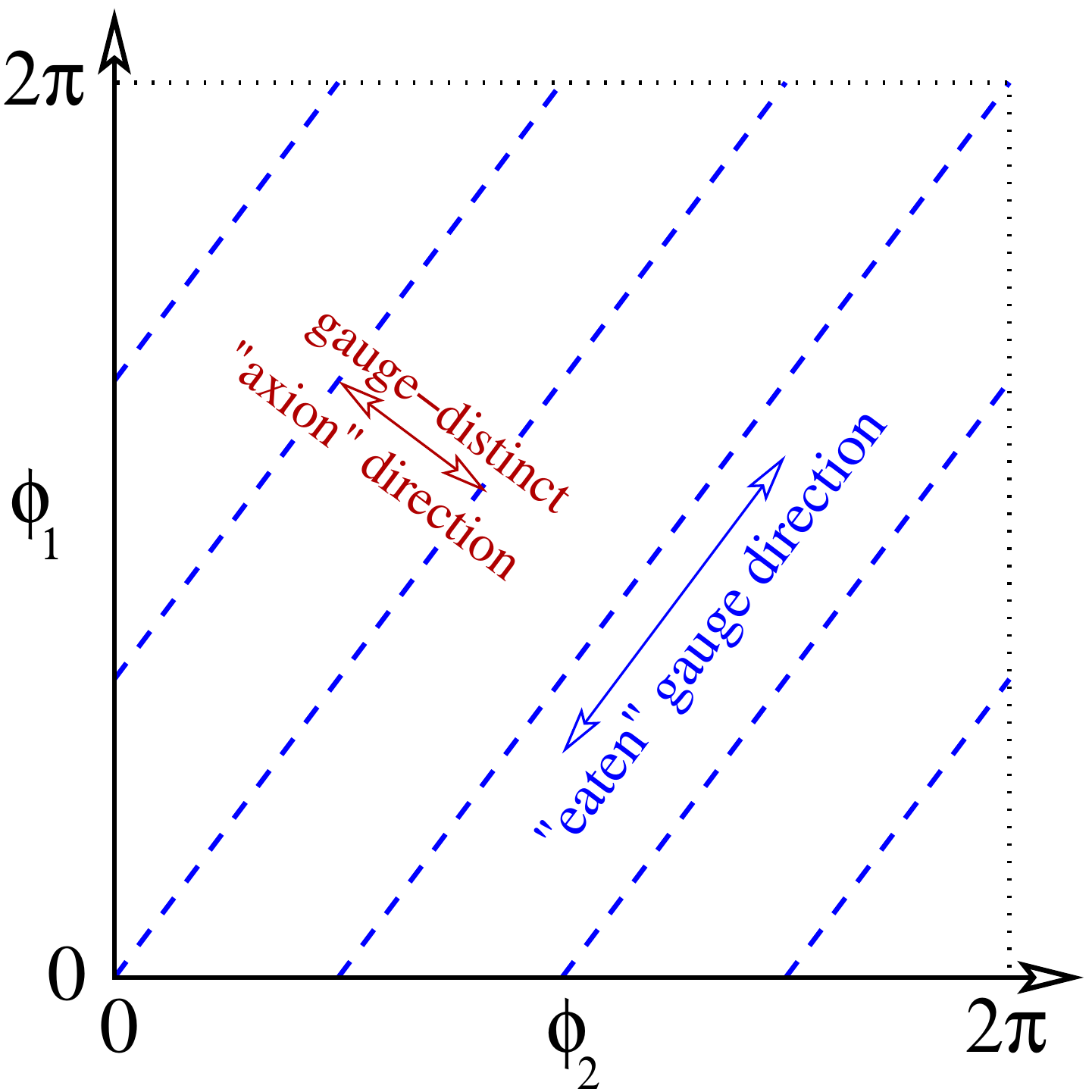}  \hfill
  $\phantom{.}$
  \caption{\label{fig:stripes}
  Space of $\varphi_1,\varphi_2$ phases $(\theta_1,\theta_2)$ for the
  case $(q_1,q_2)=(4,3)$.  The dashed (blue) line indicates phase
  pairs which are equivalent under gauge transformations.  An
  appropriate vector potential can cancel any gradient energy in the
  direction of the dashed lines.}
\end{figure}

Because $q_1$ and $q_2$ are of the same sign, the resulting
large-distance energy is smaller for the $(1,1)$ string than for the
sum of the $(1,0)$ and $(0,1)$ strings, and therefore there is an
attractive interaction between $(1,0)$ and $(0,1)$ strings, which like
to bind into a $(1,1)$ string.  Alternatively we could say that the
global charge of the $(1,0)$ string is $q_2$ and the $(0,1)$ string is
$-q_1$, so they are strongly attracted by the Goldstone-mediated
interaction and bind into a $(1,1)$ string.

\begin{figure}[htb]
  \hfill \epsfxsize=0.6\textwidth\epsfbox{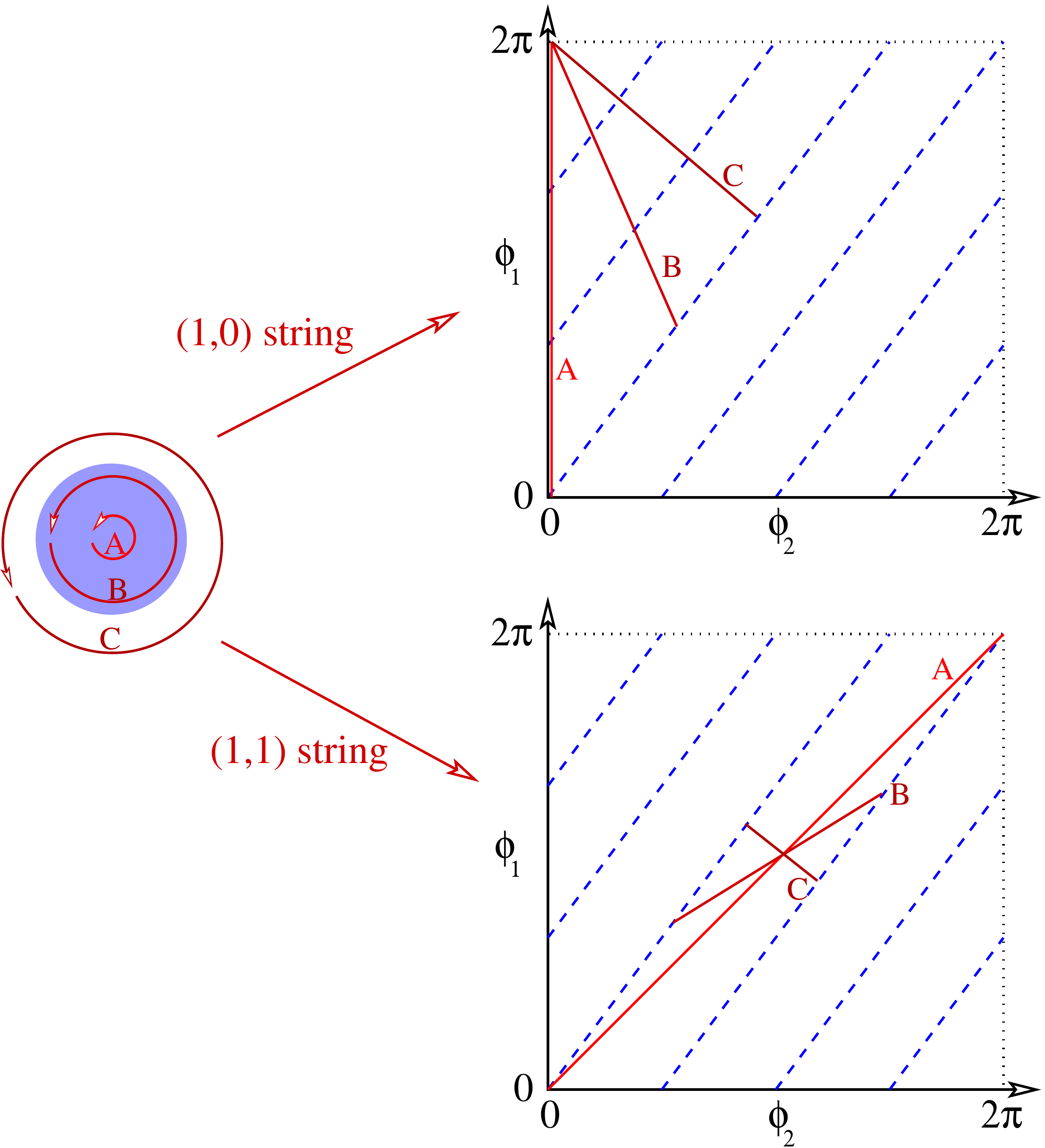}
   \hfill $\phantom{.}$
  \caption{\label{fig:strings}
  Left:  cross-section of a string, showing the magnetic field
  strength ``bundle'' and three possible loops one can take around the
  center of the string.  Right:  path through $(\phi_1,\phi_2)$
  space taken along each loop, for a $(1,0)$ string (top) and a
  $(1,1)$ string (bottom).  As more magnetic flux is enclosed, the
  component of $(\Delta \theta_1,\Delta \theta_2)$ along the
  gauge-direction is canceled, but the component in the ``global''
  direction is not.  This component is small for the $(1,1)$ string. }
\end{figure}

For a more intuitive explanation, consider Figure \ref{fig:stripes}.
It shows the set of possible phases $(\theta_1,\theta_2)$ for the two
scalar fields, in the case $(q_1,q_2) = (4,3)$.  The figure includes
a dotted line to indicate which
phase choices are gauge-equivalent.  Moving along the dotted line
corresponds to changing the gauge, or moving through space along a
gauge field; a vector potential of the right size can cancel a gradient
energy along this field direction.  The orthogonal direction, which
is unaffected by a gauge field, is the global (axion) field
direction.  A change in this direction from one blue dotted line to
the next represents a full $2\pi$ rotation in the (axial) Goldstone
direction, which explains the value of $f_a$ found in
\Eq{fa-is}.  Figure \ref{fig:strings} then shows how gradient energies
behave in a $(1,0)$ or $(1,1)$ string.  As the scalar field value
varies around a loop, the gradient energy in the dotted-blue direction
is partly canceled by $A_\phi$ gauge field.  The extent of
cancellation depends on the enclosed flux.  As one goes from a
small loop within the string core to a large loop outside the core,
more and more of the scalar gradient along the dotted-line direction is
canceled by the enclosed flux.  For the innermost loop there is no
enclosed flux, and the gradient energy is given by the distance
between the point $(\theta_1,\theta_2)=(0,0)$ to the point
$(2\pi,0)$ or $(2\pi,2\pi)$ for the $(1,0)$ or the $(1,1)$ string
respectively.  For a loop enclosing the entire flux, all gradient
energy arising from the gauge-direction is canceled.  This nearly
completely removes the gradient energy for the $(1,1)$ string, but the
reduction is modest for the $(1,0)$ string.  Therefore the $(1,1)$
string has a small gradient energy outside the core, representing a
small residual coupling to the (axion) Goldstone field, but the
$(1,0)$ string has a large gradient energy and a large coupling.

For our simulation to correspond with the global string model, it
should have Goldstone modes between a network of strings with charge-1
under the Goldstone fields.  That is, we want a network of $(1,1)$
strings only, with no other string types.  Because the $(1,1)$ string
is stable, achieving this is simply a matter of choosing the right
initial conditions.  We choose $\theta_1$ randomly and independently
at each point, $A_\mu = 0=E_i=\dot\varphi_1 = \dot\varphi_2$ initially,
and $\theta_2 = \theta_1$.  In this case only $(1,1)$ strings are
present initially, and the network never develops any other sort of
string or structure.  Since $q_2\theta_1 - q_1 \theta_2$ is initially
random, there are no long-range correlations in the Goldstone field
and the Kibble mechanism ensures a network of $(1,1)$ strings, whose
evolution should approach the scaling behavior of a network with a
tension set by the $(1,1)$ string tension and the Goldstone mode
interaction strength found above.  Our initial conditions also obey
Gauss' Law, which is then preserved by the evolution.

Now let us estimate the effective value of $\kappa$, the ratio of the
string tension to the string interaction via Goldstone modes, for such
a $(1,1)$ string.  The energy of the string's core is the energy of an
abelian Higgs string with $m_h = m_e$ and with
$f^2 = v_1^2 + v_2^2$, which is
\begin{equation}
  \label{abelian-tension}
  T_{\mathrm{abelian}} \simeq \pi ( v_1^2 + v_2^2 ) \,.
\end{equation}
The value of $\kappa$ is therefore
\begin{equation}
  \kappa = \frac{T}{\pi f_a^2}
  \simeq \frac{v_1^2 + v_2^2}
  { \frac{v_1^2 v_2^2}{q_1^2 v_1^2 + q_2^2v_2^2}}
  = \frac{(v_1^2+v_2^2)(q_1^2 v_1^2+q_2^2 v_2^2)}{v_1^2 v_2^2}
  \quad
  \longrightarrow_{v_1=v_2}
  \quad
  2(q_1^2 + q_2^2) \,.
\label{kappa-eff}
\end{equation}
This is not quite correct; the solution only coincides with the
abelian-Higgs solution for large $q_1 \gg 1$.  For finite $q_1$ we
must compute the true solution, and account for the $1/r^2$ tail of
energy arising from the long-distance Goldstone-mode content of the
string. Therefore we solve
\Eq{g-eq}, \Eq{f1-eq}, \Eq{f2-eq} numerically by multiparameter
shooting to establish the string solution and its energy.
Artificially separating the short and long distance energy
contributions by choosing $\rmin = \pi/m$ and writing the
energy-per-length stored in fields out to radius $R$ as
\begin{equation}
  \label{kappa-pi}
  T_R = \int_0^R \frac{dT}{dr} dr
  \equiv \left( \bar\kappa + \ln(mR/\pi) \right) \pi f_a^2 \,,
\end{equation}
(or equivalently,
$\bar\kappa = \lim_{R\to \infty} T_R/\pi f_a^2 - \ln mR/\pi$), we
find the values of $\kappa$ shown in Table \ref{tab1}.  The table
shows that \Eq{kappa-eff} is quite accurate, for our choice of
$\rmin$.  For future use, the table also records the small-distance
behavior of $f_1$, $f_2$, and $g$, defined as
\begin{align}
  \label{smallval}
  f_1(r) & = c_1 mr + \OO(r^3) \,, \nonumber \\
  f_2(r) & = c_2 mr + \OO(r^3) \,, \nonumber \\
  g(r) & = d (mr)^2 + \OO(r^4) \,.
\end{align}
Note that all results in the table are for the case of equal masses
and equal vacuum values; we could achieve $\bar\kappa$ values
intermediate between those shown by considering asymmetric vacuum
values $v_1 > v_2$.

\begin{table}[th]
  \centerline{\begin{tabular}{|cc|cccc|} \hline
      $q_1$ & $q_2$ & $\bar\kappa$ & $c_1$ & $c_2$ & $d$ \\ \hline
      5 & 4 &  82.290 & 0.618084 & 0.582967 & 0.0547499 \\
      4 & 3 &  50.288 & 0.621225 & 0.576319 & 0.0697357 \\
      3 & 2 &  26.283 & 0.625697 & 0.563648 & 0.0954803 \\
      2 & 1 &  10.267 & 0.630809 & 0.532107 & 0.1476050 \\ \hline
  \end{tabular}}
  \caption{\label{tab1}
    Numerical value of extra string tension $\bar\kappa$, and
    small-distance behaviors of radial functions, for several charge
    combinations.}
\end{table}

To summarize, the model of \Eq{L-2field} has an infrared description
consisting of one Goldstone mode and strings.  The strings have a
tension $\kappa \pi f_a^2$ with tunable $\kappa$ given by
\Eq{kappa-eff}.  The Goldstone phase winds by $2\pi$ in circling the
string, so the Kalb-Ramond charge of the string is correct.  This
description breaks down at a scale $m_1=m_2$, which is both the scale
setting the thickness of the string and the scale of massive
excitations off the strings.  The model can be implemented numerically
with only a little more work than the abelian-Higgs model.

\section{Numerical implementation and results}
\label{numerics}

\subsection{Numerical implementation}
\label{implement}

For an FRW spacetime in comoving coordinates and conformal time,
the action is
\begin{equation}
  \label{action}
  S = \int_0 dt \int d^3 x \; t^{k} \, \LL[\mbox{\Eq{L-2field}}]
\end{equation}
where $k$ is determined by the expansion rate; $k=2$ for radiation
(which we will study), $k=4$ for matter, \textsl{etc}.%
\footnote{%
  For an FRW universe with equation of state $P=w\varepsilon$,
  $k=4/(1+3w)$.}
Our approach is to write a spacetime-lattice discretization of the
action and to determine the update rule by extremization of this
action.  This leads automatically to a leapfrog update rule.  We use
noncompact formulation of U(1), recording gauge fields $A_i(x)$
(which ``live'' on links) directly and computing link variables
\begin{equation}
  \label{link=}
  U_i(x) = \mathrm{Pexp} \int_x^{x+a\hat{i}} -iA_i dl = e^{-ia A_i(x)}
\end{equation}
when needed.  We use an $a^2$ improved action, both for the scalar
fields and the gauge fields.  To our knowledge this has not been done
correctly before in simulating abelian Higgs fields for cosmic string
networks.  Our implementation is almost the same as a previous attempt
to $a^2$-improve the abelian Higgs mechanism \cite{Hindmarsh:2014rka},
except that we correctly modify the electric field
part, see below.  The improved scalar field ``hopping'' term is
$(4/3)$ times a nearest-neighbor term minus $(1/12)$ a next-nearest
neighbor term,
\begin{align}
  S_{\nabla \varphi} = \sum_{t=\delta n_t a}
  \sum_{x=a\vec{n}_x} & \left[ \sum_{i=1,2,3}
  t^k \left( \frac{4 | U_i^q(x,t) \varphi(x{+}a\hat{i},t)-\varphi(x,t) |^2}{3}
  \right. \right. \nonumber \\ & \left. \left.
  \phantom{\sum_{i=1,2,3} t^k} \; {}
  - \frac{|U^q_i(x,t) U^q_i(x{+}a\hat{i},t) \varphi(x{+}2a\hat{i},t)
    - \varphi(x,t) |^2}{12} \right) \right]
  \nonumber \\ &
  - (t+\delta a/2)^k \frac{|U^{q}_0(x,t) \varphi(x,t+\delta a)
    -\varphi(x,t)|^2}{\delta^2}
 \,,
\end{align}
with $q$ the charge for the specific field considered, and $\delta$
the ratio of temporal to spatial discretization; we
typically use $\delta=1/6$ which is adequate \cite{axion1}.  In
practice we fix to temporal gauge, $U_0=1$, which is numerically
convenient but not very relevant as long as we ask only gauge
invariant questions.

The noncompact magnetic field action is $(5/3)$ a square plaquette
term minus $(1/12)$ a sum on rectangular plaquettes
(the abelian version of the tree-level $a^2$ improved or Symanzik
action \cite{Weisz:1982zw,Curci:1983an}),
\begin{align}
  S_{B} = & \sum_{t,x,i>j}
  \frac{5 t^k}{6e^2} \Big( A_i(x)+A_j(x{+}a\hat{i})-A_i(x{+}a\hat{j})
  -A_j(x) \Big)^2
  \nonumber \\ & {}
  - \sum_{t,x,i\neq j}\frac{t^k}{24e^2} \Big(
  A_i(x){+}A_i(x{+}a \hat{i}) {+} A_j(x{+}2a\hat{i})
  -A_i(x{+}a[\hat{i}{+}\hat{j}]) {-} A_i(x{+}a\hat{j}) {-} A_j(x) \Big)^2
\end{align}
while the electric field action is
\begin{align}
  S_{E} = -\sum_{t,x,i} (t{+}a \delta/2)^k
  & \left[ \frac{2}{3}
    \frac{(A_i(x,t{+}\delta a)-A_i(x,t))^2}{\delta^2}
    \nonumber \right. \\ & \left. {}
    - \frac{1}{24} \frac{(A_i(x,t{+}a\delta)
      +A_i(x{+}a\hat{i},t{+}a\delta)-A_i(x,t)-A_i(x{+}a\hat{i},t))^2}
    {\delta^2} \right] \,.
\end{align}
This last modification, explained in detail in \cite{Moore:1996wn},
is necessary to make the evolution truly improved -- for instance,
without it the gauge boson dispersion relation has $a^2$ corrections.
Unfortunately it causes the $A$-field update to be implicit.  To see
this, first define $E_i(x,t) = A_i(x,t{+}a\delta)-A_i(x,t)$.  Then
$S_E$ can be written as
$\sum_{x,i} (7/12)E_i^2(x) - (1/12) E_i(x) E_i(x{+}a\hat{i})$.
Because the Lagrangian is not simply diagonal in the $E\sim \dot{A}$,
there is a difference between the time derivative of $A$ and the
canonical momentum of $A$.  It is convenient to define the quantity
$P_i(x) = -(1/12) E_i(x{-}a\hat{i})+(7/6) E_i(x) -(1/12) E_i(x{+}a\hat{i})$,
which in the continuous-time limit is the canonical momentum of the
$A$ field.  Its time-update is simple, but the relation between $P$
and $E$ must be inverted to update the $A$ field.  Because the
relation is nearly diagonal, this inversion can be done perturbatively
and proves not to be a large numerical overhead.%
\footnote{%
  This procedure can be thought of as putting a tridiagonal matrix $M$,
  with diagonal elements $[-1/12,7/6,-1/12]$, in the Hamiltonian for a
  column of $E$-fields in one direction,
  $H(E) = \sum_i EME/2$.  An alternative with a simpler update would be to
  define $N$ a tridiagonal matrix with diagonal elements
  $[+1/12,5/6,+1/12]$ and to use $H(E) = EN^{-1}E/2$.  These are
  equivalent at order $a^2$.  In the latter case we would have
  $P=N^{-1}E$ or $E=NP$.  One then stores and updates $P$, and easily
  generates $E=NP$ when needed to perform the $A$-field update.  No
  matrix inversions are required.  In each case the energy in electric
  fields is determined from $\sum_x E_i P_i/2$.  We did not implement this
  procedure, but we would use it if we were writing the code from new.}
(In Ref.~\cite{Moore:1996wn} this was the dominant cost, because the
reference works with SU(2) where $E$ is a gauge non-singlet and
parallel transportation is involved in inverting the $E$-$P$ relation.)

Improvement increases the operation count by roughly a factor of 2.5,
and most of the update effort is spent on the scalar field update.
We have implemented the resulting equations of motion in $c$ using MPI
and AVX2, obtaining $2\times 10^7$ site updates per second on an
i5core duo (two physical cores), and with MPI and AVX512, obtaining
$5\times 10^8$ site updates per second on a compute node with 2 64-core KNL
Xeon Phi processors communicating through Infiniband.  A $2048^3$
lattice fits in approximately 512G of memory.  Relative to the simple
complex-scalar model of \Eq{eq:L}, the memory demand is
$3.5\times$ as large and the compute time is $6\times$ larger on the
Xeon Phi and $16\times$ larger on the i5core.

We identify the plaquettes pierced by a string using the
gauge-invariant definition of Kajantie \textsl{et al}
\cite{Kajantie:1998bg}, applied to one of the fields (we use
$\varphi_1$).  We measure string velocity by a slight
generalization of the method of Ref.~\cite{axion1};
we use the small-$r$ expansion of $f_1(r)$:
\begin{equation}
  f_1(r) = c_1 mr +e_1 (mr)^3 + \ldots
 \,  \qquad   e_1 = -\frac{c_1(1+4 q_1 d)}{16}
  \,,
\end{equation}
and we use the string velocity estimate that near the center of a string
core, \cite{axion1}
\begin{equation}
  \gamma^2 v^2 = \frac{|\partial_t \phi|^2}{2m^2 c_1^2 v^2}
  \left( 1 - \frac{4e_1 \phi^* \phi}{c_1^2 v^2} \right)
  - \frac{4e_1(\varphi^* \partial_t \varphi+\varphi \partial_t \varphi^*)^2}
  {m^2 c_1^4 v^4}
  + \OO \left( \frac{\varphi^4 (\partial_t \varphi)^2}{m^2 v^6}
  \right) \,,
\end{equation}
which should converge to the correct velocity in the small-$a$ limit
as $(ma)^4$.%
\footnote{There are other estimators, such as that used in
  Ref.~\cite{Hindmarsh:2017qff}, which
  may be less spacing-sensitive at the spacing we consider.  It would
  be interesting to compare them systematically.  One virtue of our
  choice is that $v<1$ is manifest, since one computes $\gamma^2 v^2$
  rather than $v$ directly.}
The values of $c_1$ and $d$ are in Table \ref{tab1}.
For each plaquette pierced by a string, we average $\gamma^2 v^2$ over
the plaquette's four corners, and use this average to determine
$v,\gamma$.  Finally, we interpolate the position within a plaquette
where the string pierces it,
by fixing to the gauge where each link is $\pm 1/4$ the
value of the magnetic flux through the plaquette, and interpolating
the $\varphi_1$ field to find its zero \cite{Klaer-toappear}.
We construct strings as the series of straight segments connecting these
interpolating points \cite{Klaer-toappear}.  The overhead to identify
and record strings is a small fraction of the numerical effort.
We have compared the results using the other field $\varphi_2$ for
string identification, and consistently find string length and mean
velocity to agree within $1\%$.  We also check the average distance
between a point on the $\varphi_2$ string network and the nearest
point on the $\varphi_1$ string network:  for $q_1=4$ and $mt=512$ we
find an average distance of $0.065/m$.
Therefore each scalar describes essentially the same string network;
in particular our procedure for getting only $(1,1)$ type strings is
successful.

\subsection{Results}
\label{sec:results}

We will present some very preliminary results obtained on
$2048\times 2016\times 2000$ lattices with relatively coarse spacing
$ma=1$.  We compare the 2-scalar model at a few values of $q_1$ with
the abelian-Higgs model on the one hand, and a scalar-only global
model on the other.  In each case we take $ma = 1.0$, except for the
scalar-only model, which has thicker strings relative to the mass
($c_1$ of \Eq{smallval} is $c_1=0.412$) and which is therefore less
sensitive to the mass value.  Therefore for the scalar-only case we
used $ma=1.5$.  We have not extrapolated to finer lattice spacing,
which will in particular lead to a larger mean string velocity than
what we report below.  Our main goals are to show that the numerics
are relatively straightforward to conduct, and that many properties of
the string networks evolve smoothly from their behavior in the global
theory with low tension towards the behavior observed in local
(abelian-Higgs) networks as the string tension is increased.  But not
all at the same rate; the string velocity shifts rather quickly, while
the network density takes a much larger tension to become more
abelian-Higgs-like.  We intend to make a more comprehensive study in
the future.


We will focus on the density of the string network, the mean
string velocity, and how
``cuspy'' the strings are.  The general expectation is that the
network should evolve towards a scaling solution, where the density of
strings and other string properties scale with the system age
(see for instance \cite{Martins:1996jp,Martins:2000cs}).
We introduce the scaled network density $\xi$ and mean inter-string
distance $L_{\mathrm{sep}}$, defined as
\begin{equation}
  \label{xi-def}
  L_{\mathrm{sep}}^{-2}
  \equiv V^{-1} \int_{\mathrm{all\;string}} \gamma dl \,, \qquad
  \xi \equiv \frac{t^2}{(1+k/2)^2 L^2} \,.
\end{equation}
Here $\int \gamma dl$ is the invariant
string length, $V$ is the space volume and $t$ the time, all in
comoving conformal coordinates.  The factor $(1+k/2)^{-2}$ converts from
conformal-time based to physical-time based normalization, which is
common usage in some of the literature.
Note that different authors define the
string density in different ways, often with the same symbol.  For
instance, a recent study of abelian-Higgs networks%
\footnote{Our abelian-Higgs simulations are generally in good
  agreement with this reference, but the reference is much more
  systematic, and achieves higher statistics and better extrapolation
  towards the continuum.}
\cite{Daverio:2015nva,Hindmarsh:2017qff} uses the symbol $\xi$ to
represent the quantity we call $L_{\mathrm{sep}}$.

We will also consider the orientation autocorrelator of
the string; defining $\vec s$ as the unit tangent vector of the
string, this is defined as%
\footnote{%
  Note that we have used rest-frame string lengths without $\gamma$
  factors in this definition.  This can be improved but we have not
  yet done so.}
\begin{equation}
  \label{corr-def}
  D(\Delta l) = \frac{ \int_{\mathrm{all\;string}}
    \vec s(l) \cdot \vec s(l+\Delta l) \; dl }
  {\int_{\mathrm{all\;string}} \;  dl} ;
\end{equation}
more details will be given in a future publication
\cite{Klaer-toappear}.

Our goal is to understand the scaling behavior of string networks, and
how it depends on $\kappa$.  In practice we will never precisely
observe scaling, and in some cases we may be quite far away.
Three effects can cause scaling violations at a finite time $t$;
\begin{enumerate}
\item
  Scaling occurs when the string core has negligible size compared to
  the inter-string spacing.  Therefore it is difficult for the network
  to show good scaling at early times, before $mt \gg 1$.
\item
  The string tension does have a residual contribution from the
  Goldstone field, which increases as the string separation increases
  with time.  Since the mean inter-string spacing is expected to grow
  linearly with time, we expect
  $\kappa(t) \simeq \kappa(t_0) + \ln(t/t_0)$.
\item
  Initial conditions may start the network out as denser or thinner
  than the scaling form, and it takes time for the network to adjust.
\end{enumerate}
The severity of the first problem should scale as $(mt)^{-1}$ and
therefore becomes less severe as we achieve larger lattices which can
be run to later times%
\footnote{%
  We stop the evolution at $t=L/2$, $L$ the box length, to ensure that
  the lattice's periodicity is invisible under causal dynamics.}
(and if we can use coarser grids, $ma\sim 1$, which is why we use an
improved action).  But the severity of this problem also
depends on how \textsl{cuspy} the
strings are -- how much fine structure there is along a string.  After
all, such cusps can occur on much smaller scales than the inter-string
spacing, and we expect that any cusp structure on scales smaller than
roughly the string thickness will be lost to heavy-mode radiation,
which is unphysical from the viewpoint of the thin-string limit.
Cusps should tend to dissipate through Goldstone mode radiation.
Parametrically, because radiation involves $\pi f_a^2$ while the
energy available is set by the tension $\pi \kappa f_a^2$, string
features with length scale $l$ should dissipate in time
$t \sim l \kappa$.  Therefore smaller-$\kappa$ strings should lose
their short-scale structure, and the necessary
separation-to-core hierarchy should scale roughly linearly with
$\kappa$.  So larger-$\kappa$ networks should demand larger lattices
and later times before correct scaling sets in.

The second problem is most severe when the $\kappa$ contribution from
our abelian degrees of freedom is small; for larger $q$-values the
Abelian contribution overwhelms any small scale-dependence in the
Goldstone contribution.  Therefor this problem is mostly an issue for
purely global simulations, and perhaps for $(q_1,q_2)=(2,1)$.

The issue of initial conditions requires that, if we want to say with
any confidence that we are close to scaling, we need to see a variety
of network initial conditions, with different initial string
densities, converge to a common string density.
As stated above, we take as initial conditions that $A_i=0$ and
$\varphi_1 = v_1 e^{i\theta_1(x)}$, $\varphi_2 = v_2 e^{i\theta_1(x)}$
(same phase as $\varphi_1$) at time $t=0$ -- actually at time
$t=a$ with $a$ the lattice spacing.  This choice leads however to a
string network which starts out quite dilute compared to the scaling
solution.  Therefore we modify the evolution by setting
$k=\kstart$ until some time $\tstart$.  A large
value of $\kstart$ represents strongly overdamped evolution, leading
to slow string motion and a much denser starting network, without
excessive fluctuations.  By varying $\kstart$ and $\tstart$, we can
vary the early-time density of the string network.

\begin{figure}[htb]
  \hfill \includegraphics[width = 0.8\textwidth]{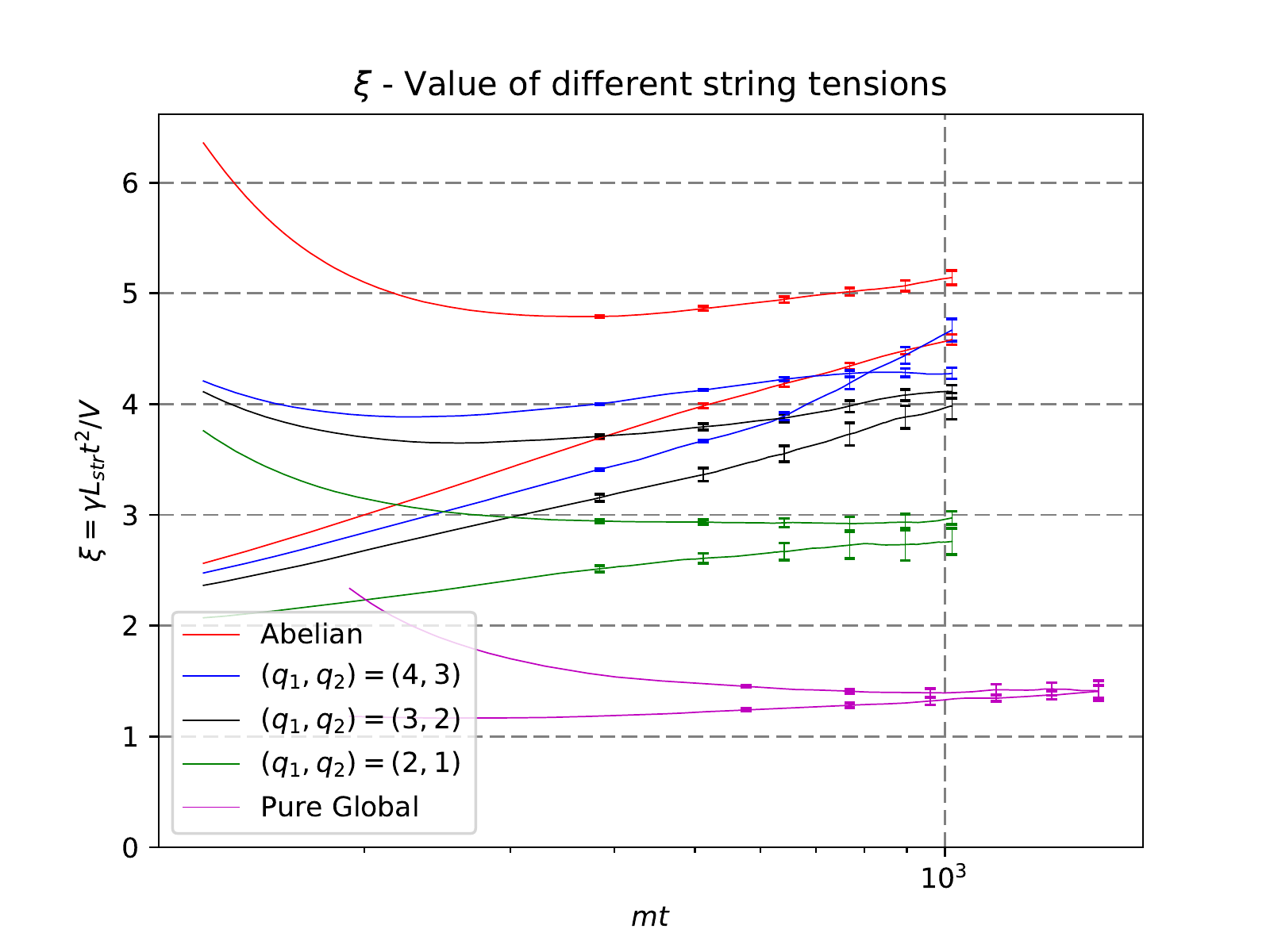}
  \hfill $\phantom{.}$
  \caption{\label{fig:length}
    Network density for different string tensions. The
    falling curves represent the overdense networks, while
    the rising curves represent the underdense
    networks.} 
\end{figure}

Figure \ref{fig:length} shows how the network density varies with the
string tension.  For each type of network, we have run two groups of
simulations, one which starts with a somewhat underdense network
($\kstart = 20$ and $\tstart =40$) and
one which starts with a somewhat overdense network
($\kstart = 50$ and $\tstart = 80$ except for the abelian-Higgs case,
where we used $\kstart = 80$ and $\tstart = 100$).
In every case the different densities converge towards each other with
time.  While many authors consider only the length of ``long''
strings, neglecting short loops, our results are based on summing over
all string lengths.  However, in all of our simulations, small loops
make up a relatively small fraction of the total string length; for
each simulation type, we find that loops satisfying
$\int \gamma dl < 2\pi L_{\mathrm{sep}}$ make up less than $10\%$ of
the total string length.  This might seem
surprising for the abelian Higgs case if one compares against
expectations from Nambu-Goto simulations
\cite{Bennett:1989yp,Allen:1990tv}
but it is consistent with recent lattice field-theory findings
\cite{Daverio:2015nva,Hindmarsh:2017qff}.  This difference, along with
the factor-2 difference in density of long strings between Nambu-Goto
simulations and lattice field-theory simulations, probably arises
because the Nambu-Goto simulations are sensitive to very short-scale
phenomena which cannot be resolved even with the largest field
theoretical simulations to date
\cite{Moore:2001px}.

The lowest density network is for pure scalar-field
simulations.  The densest network is for abelian-Higgs networks.
Increased-tension networks fall in between, with the network density
increasing as one increases the string tension.
The figure indicates that the string density increases more slowly
than linearly with $\kappa$.  For the lower-tension networks it
appears that $\xi$ converges to a good late-time value.  However for
the highest-tension networks and the abelian Higgs case, it appears
that $\xi$ continues to grow at the largest available times; indeed,
when the network is initialized as overdense, $\xi$ first falls, but
then bottoms out and rises, which clearly indicates that the dynamics
are not free of finite $1/(mt)$ corrections -- that is, we are not in
the large-time limit for these networks.  This fits with our
expectations that, the larger the value of $\kappa$, the larger the
$mt$ value must be before small-scale structure is well resolved in
the simulation.

\begin{figure}[htb]
  \hfill \includegraphics[width = 0.7\textwidth]{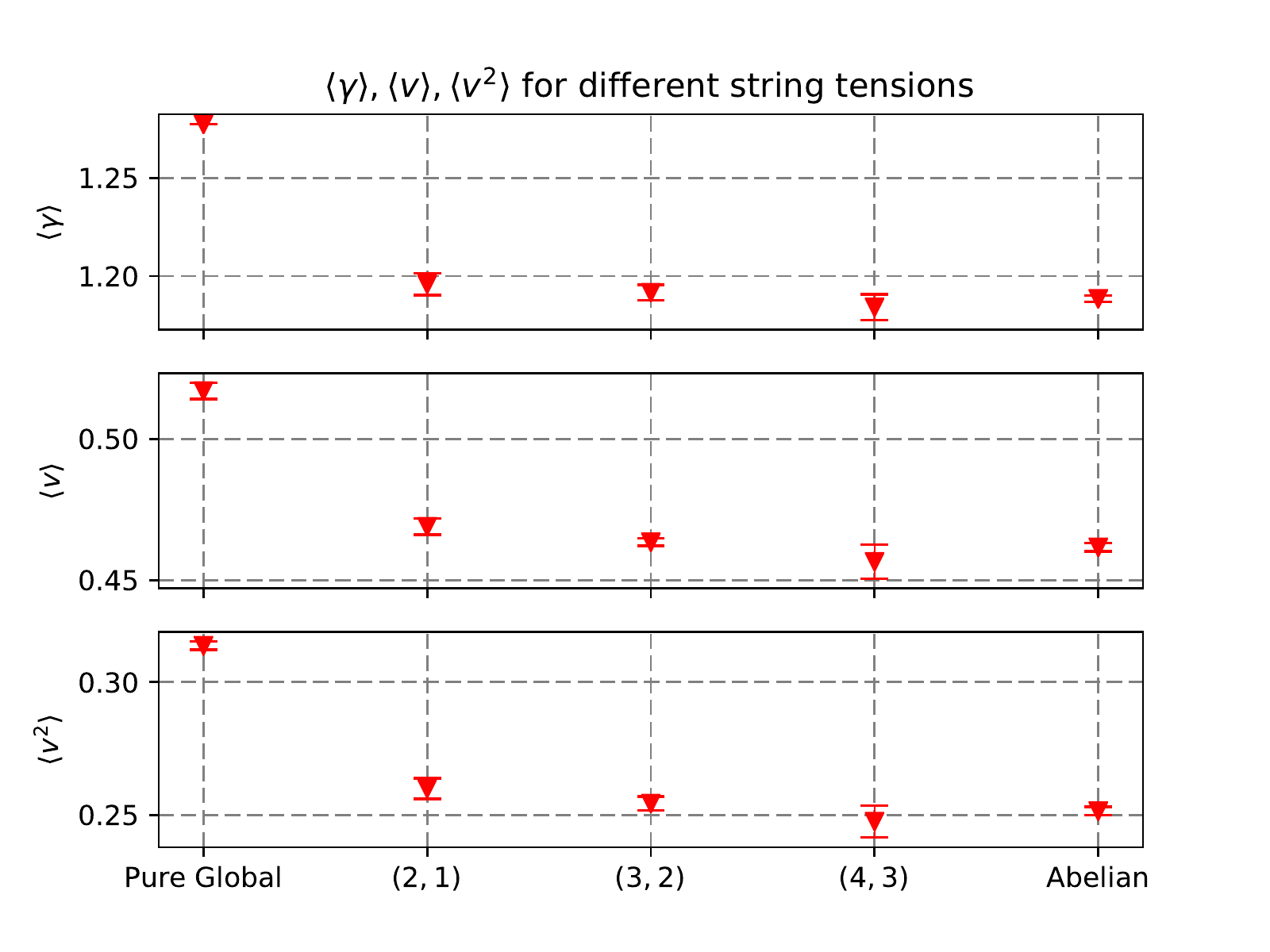}
  \hfill  $\phantom{.}$
  \caption{\label{fig:velocity}
    Mean gamma factor , velocity and squared velocity. The upper and
    lower bar indicates the mean values for the overdense network
    (upper) and for the underdense network (lower).}
\end{figure}

Figure \ref{fig:velocity} shows the mean string velocity, squared
velocity, and gamma factor, each defined as
\begin{equation}
  \left. \begin{array}{c} \langle v \rangle \\
    \langle v^2 \rangle \\ \langle \gamma \rangle \\ \end{array}
  \right\} \equiv \frac{ \int \gamma \, dl \times \left\{
    \begin{array}{c} v \\ v^2 \\ \gamma \\ \end{array} \right. }
           {\int \gamma \, dl} \,.
\end{equation}
For the overdense network the mean values are always slightly higher than for
the underdense network; the difference exceeds the statistical error
in either measurement. Therefore, rather than statistical error bars,
we have plotted the mean values of the overdense and underdense
network for the latest time we achieved, $mt=1024$.  We
stress that the velocity measurements are not extrapolated to the
continuum (in the sense of small $ma$); preliminary indications are
that all values will rise when we do so.  However the qualitative
feature, that the scalar-only theory has a higher velocity and that it
then comes down rather quickly towards the abelian-Higgs value as the
string tension is increased, appears to be robust.

\begin{figure}[htb]
  \hfill \includegraphics[width = 0.7\textwidth]{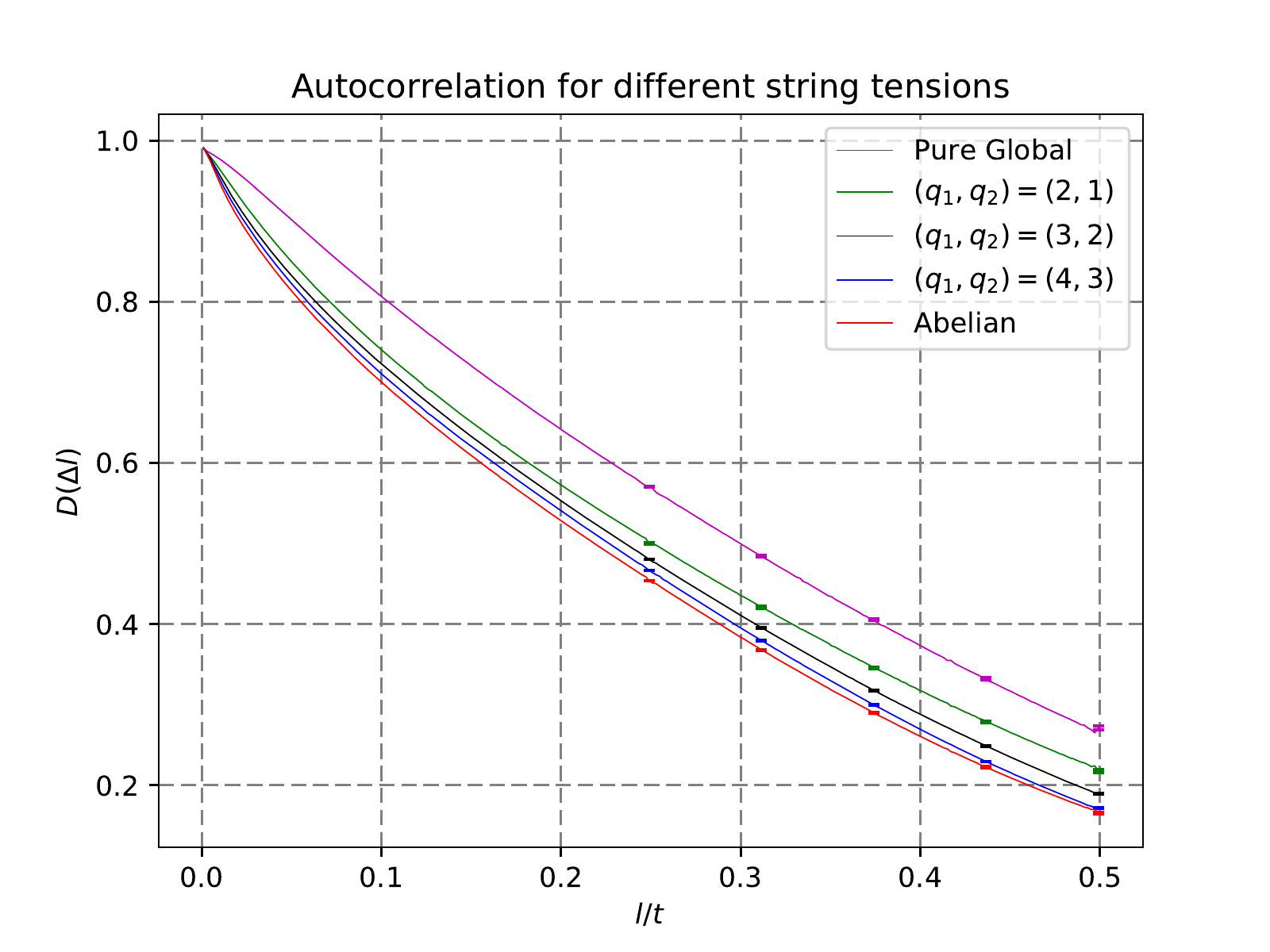}
  \hfill $\phantom{.}$
  \caption{\label{fig:angle}
    Autocorrelation of the string-direction for different string
    tensions.} 
\end{figure}

Figure \ref{fig:angle} shows the string-direction autocorrelator for
each string type.  The $x$-axis is a separation distance along a
string, normalized by the system age.  That is, an $x$-axis value of
$\ell/t=0.2$ means that we 
consider all pairs of points $(x,y)$ separated along a string by
$\int_x^y \,dl = 0.2t$.  The $y$-axis is the dot product of
their unit tangent vectors.  We see that the strings with a larger
coupling to Goldstone modes are systematically straighter (larger
correlator) than the strings with smaller or no Goldstone coupling.
The effect is especially clear at very small separation.  A string
consisting only of smooth curves would have vanishing slope at
$\ell/t=0$, while a string with perfectly sharp cusps would have a
nonzero slope at $\ell/t=0$ set by the density and angle of the
cusps.  This cuspy behavior is consistent with the abelian-Higgs
curve, but not with the scalar-only curve.  Enhanced-tension strings
lie in between, though closer to the abelian-Higgs case.  We expect
that the abelian-Higgs string and the highest-tension 2-field strings
are not yet displaying their large-$mt$ asymptotic behavior.

\section{Discussion and conclusions}
\label{discussion}

We have presented a new algorithm for simulating global string
networks, which makes it possible to consider networks with a large
value of $\kappa$, the ratio of the string tension to the coupling to
Goldstone modes.  This makes it possible to simulate
global string networks with a very large hierarchy between the Hubble
scale and the microscopic string core scale, without actually
resolving the hierarchy numerically.  Preliminary numerical studies
find that high-tension global strings behave similarly to abelian
Higgs networks, for the lattice sizes we have achieved.  In particular
we very clearly see that the density of string networks smoothly
increases from the value observed in scalar-only simulations towards
the value observed in abelian Higgs simulations, as the string tension
is increased.  Physically, we expect that the needed lattice
resolution to properly capture small scale string structure should
grow linearly with $\kappa$, meaning that large lattices are needed.
But with our approach, the lattice size need only grow as the
\textsl{logarithm} of $f_a/H$, not with $f_a/H$ itself.

Our approach has clear applications to the physics of axion production
in the early Universe.  We have shown that existing simulations must
underestimate the network density, because they fail to capture the
larger string tension.  Roughly, existing axion simulations are
comparable to the scalars-only curve in Figure \ref{fig:length}, while
the physical tension is somewhat above the $(4,3)$ line in the figure
-- which itself is probably not yet scaling, but should display a
still higher network density.  Therefore the string density in
simulations of axion production is underestimated by at least a factor
of 3.  This could certainly be important in establishing axion
production.  The large difference in string tension could also be
important. Therefore one should revisit the question of axion
production from axionic string network breakup, using our approach.
We intend to do so in the near future.  It might also be interesting
to revisit the study of the possible role of global cosmic strings in
cosmology.

\section*{Acknowledgments}

We thank the Technische Universit\"at Darmstadt and its Institut f\"ur
Kernphysik, where this work was conducted.  We thank Kari Rummukainen
and David Weir for fruitful and interesting conversations, as well as
conversations with Neil Turok, a very long time ago.  We
especially thank Mark Hindmarsh; conversations with Mark inspired this
method, although we doubt he realizes it.

\bibliographystyle{unsrt}
\bibliography{refs}

\begin{thebibliography}{10}

\bibitem{Kibble:1976sj}
T.~W.~B. Kibble.
\newblock {Topology of Cosmic Domains and Strings}.
\newblock {\em J. Phys.}, A9:1387--1398, 1976.

\bibitem{Vilenkin:1984ib}
Alexander Vilenkin.
\newblock {Cosmic Strings and Domain Walls}.
\newblock {\em Phys. Rept.}, 121:263--315, 1985.

\bibitem{Gibbons:1990gp}
G.~W. Gibbons, S.~W. Hawking, and T.~Vachaspati, editors.
\newblock {\em {The Formation and evolution of cosmic strings. Proceedings,
  Workshop, Cambridge, UK, July 3-7, 1989}}, 1990.

\bibitem{Zeldovich:1980gh}
{\relax Ya}.~B. Zeldovich.
\newblock {Cosmological fluctuations produced near a singularity}.
\newblock {\em Mon. Not. Roy. Astron. Soc.}, 192:663--667, 1980.

\bibitem{Vilenkin:1981iu}
A.~Vilenkin.
\newblock {Cosmological Density Fluctuations Produced by Vacuum Strings}.
\newblock {\em Phys. Rev. Lett.}, 46:1169--1172, 1981.
\newblock [Erratum: Phys. Rev. Lett.46,1496(1981)].

\bibitem{Kibble:1980mv}
T.~W.~B. Kibble.
\newblock {Some Implications of a Cosmological Phase Transition}.
\newblock {\em Phys. Rept.}, 67:183, 1980.

\bibitem{Ade:2013xla}
P.~A.~R. Ade et~al.
\newblock {Planck 2013 results. XXV. Searches for cosmic strings and other
  topological defects}.
\newblock {\em Astron. Astrophys.}, 571:A25, 2014.

\bibitem{Urrestilla:2011gr}
Jon Urrestilla, Neil Bevis, Mark Hindmarsh, and Martin Kunz.
\newblock {Cosmic string parameter constraints and model analysis using small
  scale Cosmic Microwave Background data}.
\newblock {\em JCAP}, 1112:021, 2011.

\bibitem{Lizarraga:2014xza}
Joanes Lizarraga, Jon Urrestilla, David Daverio, Mark Hindmarsh, Martin Kunz,
  and Andrew~R. Liddle.
\newblock {Constraining topological defects with temperature and polarization
  anisotropies}.
\newblock {\em Phys. Rev.}, D90(10):103504, 2014.

\bibitem{Lazanu:2014xxa}
Andrei Lazanu, E.~P.~S. Shellard, and Martin Landriau.
\newblock {CMB power spectrum of Nambu-Goto cosmic strings}.
\newblock {\em Phys. Rev.}, D91(8):083519, 2015.

\bibitem{Weinberg:1977ma}
Steven Weinberg.
\newblock {A New Light Boson?}
\newblock {\em Phys.Rev.Lett.}, 40:223--226, 1978.

\bibitem{Wilczek:1977pj}
Frank Wilczek.
\newblock {Problem of Strong p and t Invariance in the Presence of Instantons}.
\newblock {\em Phys.Rev.Lett.}, 40:279--282, 1978.

\bibitem{Preskill:1982cy}
John Preskill, Mark~B. Wise, and Frank Wilczek.
\newblock {Cosmology of the Invisible Axion}.
\newblock {\em Phys. Lett.}, B120:127--132, 1983.

\bibitem{Abbott:1982af}
L.~F. Abbott and P.~Sikivie.
\newblock {A Cosmological Bound on the Invisible Axion}.
\newblock {\em Phys. Lett.}, B120:133--136, 1983.

\bibitem{Dine:1982ah}
Michael Dine and Willy Fischler.
\newblock {The Not So Harmless Axion}.
\newblock {\em Phys. Lett.}, B120:137--141, 1983.

\bibitem{Visinelli:2014twa}
L.~Visinelli and P.~Gondolo.
\newblock {Axion cold dark matter in view of BICEP2 results}.
\newblock {\em Phys. Rev. Lett.}, 113:011802, 2014.

\bibitem{Davis:1986xc}
Richard~Lynn Davis.
\newblock {Cosmic Axions from Cosmic Strings}.
\newblock {\em Phys. Lett.}, B180:225, 1986.

\bibitem{Davis:1989nj}
R.~L. Davis and E.~P.~S. Shellard.
\newblock {Do Axions Need Inflation?}
\newblock {\em Nucl. Phys.}, B324:167, 1989.

\bibitem{Dabholkar:1989ju}
Atish Dabholkar and Jean~M. Quashnock.
\newblock {Pinning Down the Axion}.
\newblock {\em Nucl. Phys.}, B333:815, 1990.

\bibitem{Hagmann:1990mj}
C.~Hagmann and P.~Sikivie.
\newblock {Computer simulations of the motion and decay of global strings}.
\newblock {\em Nucl. Phys.}, B363:247--280, 1991.

\bibitem{Battye:1993jv}
R.~A. Battye and E.~P.~S. Shellard.
\newblock {Global string radiation}.
\newblock {\em Nucl. Phys.}, B423:260--304, 1994.

\bibitem{Battye:1995hw}
R.~A. Battye and E.~P.~S. Shellard.
\newblock {Radiative back reaction on global strings}.
\newblock {\em Phys. Rev.}, D53:1811--1826, 1996.

\bibitem{Chang:1998tb}
Sanghyeon Chang, C.~Hagmann, and P.~Sikivie.
\newblock {Studies of the motion and decay of axion walls bounded by strings}.
\newblock {\em Phys. Rev.}, D59:023505, 1999.

\bibitem{Yamaguchi:1998iv}
Masahide Yamaguchi, Jun'ichi Yokoyama, and M.~Kawasaki.
\newblock {Numerical analysis of formation and evolution of global strings in (
  2+1)-dimensions}.
\newblock {\em Prog. Theor. Phys.}, 100:535--545, 1998.

\bibitem{Yamaguchi:1998gx}
Masahide Yamaguchi, M.~Kawasaki, and Jun'ichi Yokoyama.
\newblock {Evolution of axionic strings and spectrum of axions radiated from
  them}.
\newblock {\em Phys. Rev. Lett.}, 82:4578--4581, 1999.

\bibitem{Yamaguchi:1999yp}
Masahide Yamaguchi.
\newblock {Scaling property of the global string in the radiation dominated
  universe}.
\newblock {\em Phys. Rev.}, D60:103511, 1999.

\bibitem{Yamaguchi:1999dy}
Masahide Yamaguchi, Jun'ichi Yokoyama, and M.~Kawasaki.
\newblock {Evolution of a global string network in a matter dominated
  universe}.
\newblock {\em Phys. Rev.}, D61:061301, 2000.

\bibitem{Hagmann:2000ja}
C.~Hagmann, Sanghyeon Chang, and P.~Sikivie.
\newblock {Axion radiation from strings}.
\newblock {\em Phys. Rev.}, D63:125018, 2001.

\bibitem{Martins:2003vd}
C.~J. A.~P. Martins, J.~N. Moore, and E.~P.~S. Shellard.
\newblock {A Unified model for vortex string network evolution}.
\newblock {\em Phys. Rev. Lett.}, 92:251601, 2004.

\bibitem{Wantz:2009it}
Olivier Wantz and E.P.S. Shellard.
\newblock {Axion Cosmology Revisited}.
\newblock {\em Phys.Rev.}, D82:123508, 2010.

\bibitem{Hiramatsu:2010yn}
Takashi Hiramatsu, Masahiro Kawasaki, and Ken'ichi Saikawa.
\newblock {Evolution of String-Wall Networks and Axionic Domain Wall Problem}.
\newblock {\em JCAP}, 1108:030, 2011.

\bibitem{Hiramatsu:2010yu}
Takashi Hiramatsu, Masahiro Kawasaki, Toyokazu Sekiguchi, Masahide Yamaguchi,
  and Jun'ichi Yokoyama.
\newblock {Improved estimation of radiated axions from cosmological axionic
  strings}.
\newblock {\em Phys.Rev.}, D83:123531, 2011.

\bibitem{Hiramatsu:2012gg}
Takashi Hiramatsu, Masahiro Kawasaki, Ken'ichi Saikawa, and Toyokazu Sekiguchi.
\newblock {Production of dark matter axions from collapse of string-wall
  systems}.
\newblock {\em Phys.Rev.}, D85:105020, 2012.

\bibitem{Hiramatsu:2012sc}
Takashi Hiramatsu, Masahiro Kawasaki, Ken'ichi Saikawa, and Toyokazu Sekiguchi.
\newblock {Axion cosmology with long-lived domain walls}.
\newblock {\em JCAP}, 1301:001, 2013.

\bibitem{Kawasaki:2014sqa}
Masahiro Kawasaki, Ken'ichi Saikawa, and Toyokazu Sekiguchi.
\newblock {Axion dark matter from topological defects}.
\newblock {\em Phys.Rev.}, D91(6):065014, 2015.

\bibitem{axion1}
Leesa Fleury and Guy~D. Moore.
\newblock {Axion dark matter: strings and their cores}.
\newblock {\em Journal of Cosmology and Astroparticle Physics}, 2016(01):004,
  2016.

\bibitem{diCortona:2015ldu}
Giovanni Grilli~di Cortona, Edward Hardy, Javier~Pardo Vega, and Giovanni
  Villadoro.
\newblock {The QCD axion, precisely}.
\newblock {\em JHEP}, 01:034, 2016.

\bibitem{Albrecht:1989mk}
Andreas Albrecht and Neil Turok.
\newblock {Evolution of Cosmic String Networks}.
\newblock {\em Phys. Rev.}, D40:973--1001, 1989.

\bibitem{Bennett:1989yp}
David~P. Bennett and Francois~R. Bouchet.
\newblock {High resolution simulations of cosmic string evolution
  evolution}.
\newblock {\em Phys. Rev.}, D41:2408, 1990.

\bibitem{Allen:1990tv}
Bruce Allen and E.~P.~S. Shellard.
\newblock {Cosmic string evolution: a numerical simulation}.
\newblock {\em Phys. Rev. Lett.}, 64:119--122, 1990.

\bibitem{Vanchurin:2005yb}
Vitaly Vanchurin, Ken Olum, and Alexander Vilenkin.
\newblock {Cosmic string scaling in flat space}.
\newblock {\em Phys. Rev.}, D72:063514, 2005.

\bibitem{Olum:2006ix}
Ken~D. Olum and Vitaly Vanchurin.
\newblock {Cosmic string loops in the expanding Universe}.
\newblock {\em Phys. Rev.}, D75:063521, 2007.

\bibitem{Hindmarsh:2017qff}
Mark Hindmarsh, Joanes Lizarraga, Jon Urrestilla, David Daverio, and Martin
  Kunz.
\newblock {Scaling from gauge and scalar radiation in Abelian Higgs string
  networks}.
\newblock 2017.

\bibitem{axion2}
Leesa~M. Fleury and Guy~D. Moore.
\newblock {Axion String Dynamics I: 2+1D}.
\newblock {\em JCAP}, 1605(05):005, 2016.

\bibitem{Goto:1971ce}
Tetsuo Goto.
\newblock {Relativistic quantum mechanics of one-dimensional mechanical
  continuum and subsidiary condition of dual resonance model}.
\newblock {\em Prog. Theor. Phys.}, 46:1560--1569, 1971.

\bibitem{Goddard:1973qh}
P.~Goddard, J.~Goldstone, C.~Rebbi, and Charles~B. Thorn.
\newblock {Quantum dynamics of a massless relativistic string}.
\newblock {\em Nucl. Phys.}, B56:109--135, 1973.

\bibitem{Nambu:1974zg}
Yoichiro Nambu.
\newblock {Strings, Monopoles and Gauge Fields}.
\newblock {\em Phys. Rev.}, D10:4262, 1974.

\bibitem{Kalb:1974yc}
Michael Kalb and Pierre Ramond.
\newblock {Classical direct interstring action}.
\newblock {\em Phys. Rev.}, D9:2273--2284, 1974.

\bibitem{Vilenkin:1986ku}
Alexander Vilenkin and Tanmay Vachaspati.
\newblock {Radiation of Goldstone Bosons From Cosmic Strings}.
\newblock {\em Phys. Rev.}, D35:1138, 1987.

\bibitem{Hindmarsh:2014rka}
Mark Hindmarsh, Kari Rummukainen, Tuomas V.~I. Tenkanen, and David~J. Weir.
\newblock {Improving cosmic string network simulations}.
\newblock {\em Phys. Rev.}, D90(4):043539, 2014.
\newblock [Erratum: Phys. Rev.D94,no.8,089902(2016)].

\bibitem{Weisz:1982zw}
P.~Weisz.
\newblock {Continuum Limit Improved Lattice Action for Pure Yang-Mills Theory.
  1.}
\newblock {\em Nucl. Phys.}, B212:1--17, 1983.

\bibitem{Curci:1983an}
G.~Curci, P.~Menotti, and G.~Paffuti.
\newblock {Symanzik's Improved Lagrangian for Lattice Gauge Theory}.
\newblock {\em Phys. Lett.}, 130B:205, 1983.
\newblock [Erratum: Phys. Lett.135B,516(1984)].

\bibitem{Moore:1996wn}
Guy~D. Moore.
\newblock {Improved Hamiltonian for Minkowski Yang-Mills theory}.
\newblock {\em Nucl. Phys.}, B480:689--728, 1996.

\bibitem{Kajantie:1998bg}
K.~Kajantie, M.~Karjalainen, M.~Laine, J.~Peisa, and A.~Rajantie.
\newblock {Thermodynamics of gauge invariant U(1) vortices from lattice Monte
  Carlo simulations}.
\newblock {\em Phys. Lett.}, B428:334--341, 1998.

\bibitem{Klaer-toappear}
Vincent~B. Klaer and Guy~D. Moore.
\newblock {Global string networks and the one-scale model} (in preparation).

\bibitem{Martins:1996jp}
C.~J. A.~P. Martins and E.~P.~S. Shellard.
\newblock {Quantitative string evolution}.
\newblock {\em Phys. Rev.}, D54:2535--2556, 1996.

\bibitem{Martins:2000cs}
C.~J. A.~P. Martins and E.~P.~S. Shellard.
\newblock {Extending the velocity dependent one scale string evolution model}.
\newblock {\em Phys. Rev.}, D65:043514, 2002.

\bibitem{Daverio:2015nva}
David Daverio, Mark Hindmarsh, Martin Kunz, Joanes Lizarraga, and Jon
  Urrestilla.
\newblock {Energy-momentum correlations for Abelian Higgs cosmic strings}.
\newblock {\em Phys. Rev.}, D93(8):085014, 2016.
\newblock [Erratum: Phys. Rev.D95,no.4,049903(2017)].

\bibitem{Moore:2001px}
J.~N. Moore, E.~P.~S. Shellard, and C.~J. A.~P. Martins.
\newblock {On the evolution of Abelian-Higgs string networks}.
\newblock {\em Phys. Rev.}, D65:023503, 2002.

\end{thebibliography}

\end{document}